\documentstyle[12pt]{article}          
\catcode`\@=11
\@addtoreset{equation}{section}

	\newdimen\eqskip
	\newdimen\txtskip
	\baselineskip=\txtskip
\def	\be		{\begin{equation}}
\def	\ee		{\end{equation}}
\def	\ba		{\begin{eqnarray}}
\def	\ea		{\end{eqnarray}}

\def	\=		{\;=\;}
\def	\frac		#1#2{{#1 \over #2}}

\def	\to		{\rightarrow }

\def	\jpsi		{\mbox{$J/\psi$}}
\def	\psitwos	{\mbox{$\psi(2S)$}}

\def	\b		{\mbox{$b$}}

\def	\muf		{\mbox{$\mu$}}
\def	\muo		{\mbox{$\mu_0$}}

\def    \pt	        {\mbox{$p_t$}}
\def    \ptmin	        {\mbox{$p_t^{min}$}}

\def    \et	        {\mbox{$E_t$}}

\def \mb   {\mbox{$m_b$}}

\def \mtr  {\mbox{$m_T$}}

\def \lqcd  {\mbox{$\Lambda_{QCD}$}}
\def \lzero {\mbox{$\Lambda_4^0$}}
\def \lplus {\mbox{$\Lambda_{4}^+$}}
\def \lff   {\mbox{$\Lambda_{4}$}}

\begin{document}
\begin{titlepage}
\nopagebreak
\begin{flushright}
IFUP-TH 2/93 \\
hep-ph/9302279 \\
January 1993
\end{flushright}
\vfill
\begin{center}
{\Large { \bf \sc On the $B$ and $J/\psi$ Cross Section Measurements \\[1cm]
at UA1 and CDF}}
\vfill
{\it \large Michelangelo L. Mangano}
\vskip .3cm
{INFN, Scuola Normale Superiore and Dipartimento di Fisica, Pisa, Italy}\\
\vskip .6cm
\end{center}
\nopagebreak
\vfill
\begin{abstract}
{We analise the implications of the measurement of $B$ and $J/\psi$ inclusive
\pt\ distributions performed in $p\bar p$ collisions by the UA1 and CDF
experiments. }
\end{abstract}
\vfill
\end{titlepage}
%
%
\section{Introduction}
Heavy quark production in high energy hadronic collisions consitutes a
fundamental arena for the study of perturbative QCD \cite{nason}.
The comparison of
experimental data with the predictions of QCD provides a necessary check that
the ingredients entering the evaluation of hadronic processes (partonic
distribution functions and higher order corrections) are under control and can
be used to evaluate the rates for more exotic phenomena or to extrapolate the
calculations to even higher energies. The estimates of production rates for the
elusive {\em top} quark rely on the understanding of heavy quark
production properties within QCD.
Following the initial encouraging agreement between the
UA1 measurements \cite{ua1_psi1,ua1_b1}\ of inclusive bottom-quark and
\jpsi\ \pt\ distributions and the best available QCD calculations
\cite{nde,nde_pt,gms}, the subject has acquired a new particular interest after
the latest measurements by UA1 \cite{ua1_psi2,ua1_b2} and CDF
\cite{cdf_psiK,cdf_psi,cdf_dpf} and after the recent studies of the small-$x$
behaviour of hadro-production cross-sections \cite{smallx1,smallx2,smallx3}.

On the experimental side UA1 has confirmed with the most recent analyses of the
\b\ sample the agreement between next-to-leading order (NLO) QCD and data,
extending it to large values of \pt \cite{ua1_b2}. At the same time, however,
UA1 has found a significant disagreement between its latest \jpsi\ \pt\
spectrum \cite{ua1_psi2}\ and the available leading order (LO) QCD
calculations.
Independently, CDF preliminary and published results indicate a clear
discrepancy with respect to theoretical expectations for normalization and
shape of the \b\ and \jpsi\ \pt\ spectra \cite{cdf_psiK,cdf_psi,cdf_dpf}.

On the theoretical side, studies have indicated that large corrections to
the NLO evaluation of the \b\ cross section should be expected at very large
energies \cite{smallx1,smallx2,smallx3}.
A precise quantitative definition of how large these
energies should be for these effects to become dominant is however still
missing, and the question of whether these effects can indeed justify the
findings by CDF is still open.

In parallel, several measurements \cite{NMC,CCFR}\ have improved the knowledge
of the parton distribution functions (PDF) down to values of $x$ of the order
of 0.01, and the relative new sets of parametrizations have now become
available \cite{newmrs,CTEQ}.

Both these latest developments could have some bearing on the measurements of
the \b\ and \jpsi\ production rates at UA1 and CDF, because the values of $x$
probed by CDF are approximately a factor of 3 smaller than
those explored by UA1. Therefore CDF is more sensitive than UA1 to the
possible uncertainties of the extrapolation to small-$x$ of both PDF's and
partonic cross-sections.

Attempts have also been made \cite{berger1,berger2} to incorporate directly
into the fits of gluon distributions the experimental information contained in
the $b$ cross section measurements by UA1 and CDF. However when the new PDF
measurements quoted above are included as an additional constraint, there seems
not to be enough freedom to correct the theoretical prediction by the needed
amount to produce a fully satisfactory agreement with CDF data.

In this letter we will reconsider these data and use the most recent inputs to
shed, if possible, some more light on these problems.

\section{The inputs of the theoretical calculations}
We will start by discussing the ingredients of the calculation presented here
and by comparing them to the analyses quoted previously.

\jpsi\ production is evaluated by summing the result of inclusive \b\
production followed by $B\to \jpsi +X$ decays, direct \jpsi\ production and
production via radiative decays of $\chi$ states. These last two processes we
will refer to as {\em direct charmonium production}, while the first one will
be
referred to as \b-decays. The matrix elements for charmonium production are
taken from the original calculations of ref~\cite{onia}, and therefore coincide
with those used in previous studies (\cite{gms}, from now on indicated by GMS).

As an improvement with respect to GMS we will use:
\begin{enumerate}
\item the full NLO matrix elements for the production of \b\ quarks before
their decay into \jpsi;
\item the most recent inclusive $B\to \jpsi +X$ and $B\to \psi(2S)+X$
fragmentation spectra \cite{btopsi,cdf_psi};
\item the most recent parametrizations of PDF described in \cite{newmrs},
exploring the dependence on the value of the 2-loop \lff\ within the one
standard deviation range $\lff=215\pm60 \mbox{MeV}$.
\end{enumerate}
This third point, in particular, will cause a significant drop in the
prediction of the absolute rates for charmonium compared to GMS, because the
PDF's used in GMS had a much larger value of \lff\
($\Lambda^{1-loop}_4=400$ MeV versus $\Lambda^{2-loop}_4=215$ MeV as the
central value of the new sets, corresponding to $\Lambda^{1-loop}_4=140$ MeV.).

As was done in GMS we will smear the \b-quark \pt\ with a Peterson
fragmentation function \cite{peterson}\ before its decay into $\psi$'s:
\be
	\frac{dN}{dz} \propto \frac{z(1-z)^2}{[\epsilon z + (1-z)^2]^2},
\ee
with $\epsilon=0.006$. Furthermore we will assign to the \b-meson after
fragmentation a mass of 5.28 GeV, regardless of the input \b-quark mass.

The central values we use for the \b-quark mass and the
factorization/renormalization scale \muf\ are \mb=4.75 GeV and
$\muf^2=\muo^2=\pt^2+\mb^2$ \footnote{Notice that this prescription is slightly
different from the one chosen in \cite{nde_pt}, where
$\muf^2=\pt_{min}^2+\mb^2=\mtr^2$ was used to estimate $\sigma(\pt>\pt^{min})$.
This difference amounts to an effect of the order of $-20$\% at the smallest
values of \pt.}. We will consider in the following the effect of changing these
parameters. The numerical values of other parameters used in the calculations
are contained in Tables~\ref{tparam} and \ref{tbr}.

\subsection{Discussion of the uncertainties in the calculations}
Several factors contribute to the uncertainty of the calculations. We will
discuss first those intrinsic to the perturbative approach and will consider
later those related to external inputs such as structure functions, \lqcd\ or
potential models in the case of the charmonium production.

First we consider \b\ production. As is well known, the radiative
corrections to \b\ production are rather large and extremely unstable under
changes in factorization scale \muf, in particular at the larger CM energies.
The standard way to establish how reliable the perturbative expansion is, is to
vary the factorization scale \muf\ within some range of the order of the hard
energy scale relevant for the process considered. If the observed change in
cross sections is large, one might think of selecting a best value for \muf\ by
fitting the measured rates. Since the dependence on \muf\ varies as a function
of the beam energy (indicating that the effect of yet higher order corrections
is not the same at different energies)  there is no reason a priori for which
the expression for \muf\  fitted at one value of the beam energy should be the
same at different energies.

Therefore the uncertainty related to the choice of \muf\ cannot be removed by
performing independent measurements at different energies.

Likewise, there is no guidance on what is the proper range within which to
allow \muf\ to vary. If \muo\ is the typical energy scale of a given process
(say the transverse mass of a heavy quark or the \et\ of a jet), it is
customary to vary \muo/2 $< \muf < $2\muo. There are several indications,
however, that when working at a fixed order in perturbation theory scales
significantly smaller than the {\em natural} scale are needed to reproduce the
data. As examples, we quote the cone-size dependence of the jet \et\
distributions in hadronic collisions \cite{cdf_jet_shape} or the jet
multiplicity distributions in $e^+e^-$ \cite{lep}.
Wherever all-order calculations have
become available, these indicate that the resummation of leading and
sub-leading large logarithmic terms at any order in the perturbative expansion
restores the insensitivity to \muf\ and allows \muf\ to be chosen of the order
of the natural scale \muo\ \cite{webber}.

We therefore believe it is legitimate to push \muf\ to values as low as
possible, compatibly with the range allowed by the PDF parametrizations.  In
our case, we will consider the range \mtr/4 $< \muf < $\mtr, which will give us
$\muf^2>5$ GeV$^2$ for all values of \pt\ probed by the CDF data, and therefore
does not include regions of $Q^2$ which are not under the control of the DIS
data.

Similar considerations apply to the case of direct quarkonium production of
\jpsi's. Here the situation is even worse, because only the LO
production processes are available and the expected \muf\ dependence is even
more significant. In GMS the effect of possible higher order terms was
parametrized in terms of a constant $K$ factor, chosen to have a value of 2 to
reproduce ISR data. As mentioned above, however, there is no reason a priori
why this same value for $K$ should apply at the significantly higher energies
used at UA1 and CDF. Once again, therefore, we will choose to probe the
possible effects of higher order terms by selecting different values of \muf.
In accordance with the choice made for the \b\ production, we will choose
\mtr/4 $< \muf < $\mtr, where in this case \mtr\ represents the transverse mass
of the
quarkonium state. As will be shown later on, different values of \muf\ will not
only change the absolute normalization but will also affect the shape of the
\pt\ distribution of \jpsi's.

In addition to the above uncertainties, one should add the uncertainty in the
evaluation of the parameters of the quarkonium states entering the estimate of
their production cross section. As an example, we quote a recent study
\cite{bodwin}\ of  $\vert R'(0) \vert^2$ -- the first derivative of the wave
function at the origin for P-wave states --  indicating a value for $\chi_c$
states which is approximately 50\% higher than what obtained from previous
models \cite{hagiwara}.  The values used in this paper are given in Table~1,
and follow the old results of \cite{hagiwara}.

\subsection{Numerical Results}
We collect the results obtained for the \b\ and \jpsi\ cross sections as a
function of the various input parameters in a series of tables. As a standard
reference we will use sets D0 and D-- of the recent MRS PDF parametrization
\cite{newmrs}.  We will use two values of \lqcd, \lzero=215 MeV and \lplus=275
MeV, corresponding respectively to the central value and to one standard
deviation
above the central value obtained from the fit. Tables~\ref{tbcdf1}\
and \ref{tbcdf2}\ contain the bottom quark \pt\ distribution
integrated above a given \pt\ at 1.8 TeV and for the two extreme values of
\muf, \muf=\mtr\ and \muf=\mtr/4. The quark is required to satisfy $\vert y
\vert <1$, to allow comparison with the CDF data. Tables~\ref{tbua11}\ and
\ref{tbua12} contain the same information, but at 630 GeV and with $\vert y
\vert < 1.5$, to allow comparison with UA1 data.

Several comments are in order. First of all notice that while the use of the
more singular set of structure functions leads to larger values of the
total cross sections ($\pt>0$) at 1.8 TeV, the opposite happens at 630 GeV.
This is because the higher density of gluons at small $x$ described by set D--
forces via momentum sum rules a depletion at larger values of $x$. Since UA1 is
sensitive to larger values of $x$, an overall decrease in the total cross
section is observed.

Notice also, on the other hand, that even at CDF the singular gluon
parametrization D-- will give a cross section smaller than the set D0 as soon
as we consider transverse momenta of the \b\ above 10 GeV -- which is the
region
where most of the CDF data are.
Since above 10 GeV the shapes of the integrated \pt\ distributions for the two
parametrization D0 and D-- are similar, this indicates that the measurement of
the cross section for \b\ production in this region cannot be reliably used to
extract via extrapolation to \pt=0 the total \b\ production cross section. For
example, while the region $\pt>10$ GeV represents 10\% of the total cross
section according to D0 and using \lqcd=215 MeV,  the same region represents
only 7\% of the total according to D-- and using \lqcd=275 MeV.
A similar exercise at the UA1 energy indicates a more reliable extrapolation.

As already indicated in \cite{nde_pt}, the dependence on the value of the \b\
mass is not significant. In Table \ref{tbmass}\ we show a comparison between
the integrated \b\ \pt\ distribution  obtained using \mb=4.5 and \mb=4.75 GeV.
The difference is of the order of 20\% for the total cross section, but becomes
negligible for \pt$>$ 10 GeV.

In Table~\ref{tpsicdf}\ we present the integrated \pt\ distribution of \jpsi\
mesons, calculated at CDF energy and divided into the direct quarkonium and $B$
decay contributions. The relative fraction due to $B$ decays, indicated by
$f_B$, is also shown as a function of the \pt\ threshold, and the dependence on
\lqcd\ and \muf\ is studied by considering the central case of
\lqcd=215 MeV, \muf=\mtr\ and the extreme case of \lqcd=275 MeV, \muf=\mtr/4.
A priori there is no reason why the same factorization scale should be used for
the two contributions, as the two physics processes are entirely different.
Furthermore the $B$ decay is evaluated at NLO, while as mentioned previously
only the LO terms are available and included in the quarkonium term.
Nevertheless we take here the value of $\mu$ for the two processes to be the
same,in order to extract and indicative range of values for $f_B$. The value of
$f_B$ plays an important role in the experimental determination of the $B$
cross section out of the measurement of the inclusive \jpsi\ rate, and the
range of values exhibited by the tables indicates what is the systematic
uncertainty that one should expect in deriving $f_B$ from the theory.  Parallel
results for UA1 energy are shown in Table~\ref{tpsiua1}.

The most important thing to notice about these tables is the fact that
the $B$ contribution only changes within a factor of 2 by changing \muf, while
a variation ranging from a factor of 7 to 10, depending on \pt, is observed for
the charmonium case.
This indicates that the LO prediction for direct charmonium is very poor, and
very large NLO corrections should be expected.

\section{Comparison with the data and discussion}
We will now compare UA1 and CDF data with the results obtained
so far. For these comparisons we use the results obtained with the D0 PDF set
and with the two different choices $(\muf,\lqcd)=(\mtr,215MeV)$ and
$(\muf,\lqcd)=(\mtr/4,275MeV)$, which provide an acceptable upper and lower
limit to the band of current theoretical uncertainty relative to the inputs
discussed above.

Figs. \ref{fua1_bpt}\ and \ref{fua1_Bmespt}\ show the UA1 measurement of the
inclusive \b-quark and $B$ meson \pt\ distribution integrated above a given
threshold \ptmin\ and with $\vert y_{b,B}\vert < 1.5$.
The agreement between data and theory
observed in ref.\cite{ua1_b2}\ is confirmed, even though the central value of
the prediction has dropped by almost 50\% as a consequence of the smaller value
of \lff\ in the central  MRS fit compared to the central DFLM fit (215 versus
260 MeV). Also the $B$-meson spectrum is well consistent with what expected
from a Peterson fragmentation model, as anticipated in \cite{nason}. Notice,
however, that a priori there is no guarantee that the Peterson model should
work for values of \pt\ of the order of the $B$ mass, as in this region
corrections to factorization could be significant.

Fig.\ref{fua1_psi}\ shows the inclusive \pt\ differential distribution for
\jpsi's produced with $\pt>5$ GeV and $\vert y \vert < 2$. We superimpose
the contributions from direct charmonium production, b-decays and the sum of
the two.  The data fall all inside the theoretical band. Since as mentioned
above there is no reason to expect the expression for \muf\ to be the same for
the two contributions -- while for the sake of simplicity we imposed this in
adding the separate terms in the figure -- a better fit to the data could be
obtained by choosing \muf=\mtr/4 for the charmonium and \muf=\mtr\ for the B
production.

Fig.~\ref{fcdf_bpt}\ shows the integrated \pt\ distribution of \b\ quarks with
$\vert y \vert < 1$ from CDF, compared to the results of the NLO calculation.
The data are taken from published results as well as from recent public
presentations \cite{cdf_psiK,cdf_psi,cdf_dpf}. As already observed in
\cite{cdf_psiK}\ there is a clear excess in the observed rate at small \pt.
At larger values of \pt, in the region of the inclusive $b\to l+X$
measurements, the data are consistent with the upper estreme of the theoretical
band. A similar feature is observed in the $\psi$ differential \pt\
measurement, shown in fig.~\ref{fcdf_psi}.

Equally worrysome is the comparison between theory and data in the
case of the \pt\ spectrum of the \psitwos, shown in fig.~\ref{fcdf_psi2pt}.
As noted in \cite{cdf_psi,gms}, the expected contribution from direct
quarkonium
production is heavily suppressed. We confirm this estimate, and verified that
it
remains true even allowing for the variation of \muf\ within the
\muo/4$<\muf<$\muo\ range.

Is it possible to explain the patterns observed by UA1 and
CDF in a unified fashion by invoking generic small-x effects, either from PDF's
or from violation of factorization? Rather than studying this question by
directly attempting to modify the gluon densities, as done in
refs.\cite{berger1,berger2}, we will address it here by considering the
following quantity:
\be
	\sigma(x_g<x;\pt^b>\pt^{min}) \=
	\int_0^x \; dx_g \, \frac{d\sigma(\pt^b>\pt^{min})}{dx_g},
\ee
namely the contribution to the integrated \pt\ distribution coming from partons
with momentum fraction smaller than a given value of $x$.  We plot this
variable as a function of $x$ and for different values of $\pt^{min}(b)$  in
Fig.~\ref{fxua1}\ for UA1 and CDF. We only integrated over \b\ quarks  within
the regions of acceptance of the experiments, namely  $\vert y_b\vert <1.5$ for
UA1 and $\vert y_b\vert <1$ for CDF. Since the contribution to the
cross-sections due to the $q\bar q$ and $qg$ initial states are negligible for
the relevant regions of \pt\ we are concerned with, we limited ourselves to the
$gg$ process and normalized the curves to the value of 1 at $x=1$. Therefore
the plotted functions represent the fraction of  cross-section due to gluons
with $x_g<x$.

The first thing to notice is that the distribution corresponding to $\pt>5$ GeV
at UA1 lies between the curves for $\pt>10$ and $\pt>20$ GeV at CDF,
consistently with the factor of 3 difference in beam energy. The second thing
to notice is that at CDF energies the contribution to the cross section for
$\pt>10$ GeV from the region $x<0.01$ is less than 20\%. Furthermore no
contribution at all comes from the region $x<0.003$. We verified that
different fits of the NMC and CCFR data, obtained in Ref.~\cite{CTEQ}, give
gluon densities which differ, over the relevant kinematic range,
by no more than 10\% from the MRSD0 set used here.
Since all of these gluon parametrizations do not differ significantly from
previous extrapolations, we conclude that the knowledge of the gluon density in
the relevant region $0.1>x>0.01$ and $Q>5$ is today rather solid.
We therefore expect that only dramatic changes in the guon densities in the
region  $0.003<x<0.01$ will lead to a change of
a factor of 2 in the cross section integrated above \pt=10 GeV.

Therefore while it is tempting to conjecture that the ignorance about the
behaviour of the gluon densities at small-$x$ could explain the discrepancy
between the overall rates measured by UA1 and CDF and the difference in slope
of the CDF spectra compared to theory, we find no evidence that this assumption
is justified. Rather, we find that the region $x_g<0.01$ is marginal in
the production of \b\ quarks or $\psi$'s passing the required acceptance and
\pt\ cuts imposed by the two experiments.

The effect of the small-x corrections to the partonic cross-section considered
in \cite{smallx1,smallx3}\ is more difficult to estimate. In fact these
phenomena alter the kinematic connection between \pt\ and $x$, since they
predict that initial state gluons with a given momentum fraction   $x$ can have
a \pt\ non negligible w.r.t. $xE_{beam}$. This is equivalent to having an
intrinsic \pt\ of the order of the scale of the hard process itself, namely
\mb. As a result, the region with $x_g<0.01$ could provide a significant
contribution to the rate for $p_t^b>10$ GeV, thanks to the transverse momentum
smearing induced by this sort of {\em small-x primordial \pt}. Even though it
was found in ref.\cite{smallx1}\ that these small-$x$ effects can add at most
50\% of the NLO contribution to the total \b\ cross section at 1.8 TeV, no
explicit indication is given on the \pt distribution of this additional 50\%.
Since the cross section observed experimentally ($\pt_b > 8.5 $ GeV) represents
of the order of 10\% of the total rate at NLO, we cannot exclude that the \pt\
smearing induced by these effects be responsible for the factor of 2-3
discrepancy observed between data and NLO predictions. Notice that the
hypotesis of a \pt\ smearing would help understanding not just the
rate deficiency, but also the apparent difference in shape between NLO and
data.
A quantitative statement regarding these  possiblities will only come from more
explicit studies along the lines of ref.\cite{smallx3}.

While we await for more explicit calculations, it might be worth exploring some
additional consequences of this scenario. In addition to trying to push the
measurement of the $b$ cross section to even smaller values of \pt, it would be
important to study correlations between the pair of $b$ quarks. NLO
calculations exist for these correlations \cite{mnr}. If the small-$x$ effects
were to behave as indicated previously, we would expect to observe a flattening
of the $\Delta\phi$ and $\pt^{b\bar b}$ distributions w.r.t the NLO prediction.
Here $\Delta\phi$ represents the difference in azimuth between the $b$ and the
$\bar b$, and $\pt^{b\bar b}$ represents the transverse momentum of the pair.
The flattening would be caused by the additional intrinsic \pt\ due to the
gluon transverse momentum.

Measurements of the $\Delta\phi$ correlations have been performed by UA1
\cite{geiser}, indicating a good agreement with the NLO calculation \cite{mnr}.
This result does not resolve the issue, though, because the agreement of the
NLO $b$ cross section with the data suggests that the energy at UA1 is below
the threshold for the possible onset of these new small-$x$ phenomena.

\section{Conclusions}
After allowing for rather generous estimates of the theoretical uncertainties
involved in the calculations currently available for \b\ and \jpsi\ production
in hadronic collisions, we conclude that the most worrisome points of
discrepancy can be summarised as follows:
\begin{enumerate}
\item The production of direct charmonium both at CDF and UA1 is much
more abundant than would be obtained from the LO calculation using a standard
value of \muf=\mtr. The \jpsi\ can be explained by using \muf=\mtr/4, which
however gives a rate 8-10 times larger than for \muf=\mtr, indicating a rather
unstable perturbative expansion. However this is not sufficient to explain the
rate of \psitwos\ production.
\item \b\ production at CDF for values of \pt\ around 10 GeV is significantly
larger than can be accomodated by current estimates of the higher order effects
or by possible structure function uncertainties. Using the extreme value of
\muf=\mtr/4 is not sufficient to explain all the data points, the discrepancy
being still larger than a factor of 2. We cannot however exclude that the
solution be in the large  smearing induced by a small-$x$ intrinsic \pt\ of the
initial state gluons. At larger values of \pt\ we believe that the consistency
between data and theory is acceptable.
\end{enumerate}

What other effects could be responsible for the
remaining discrepancies? It should be noticed that the two points above might
not be
uncorrelated. In fact the absolute normalization of the two CDF points at lower
\pt, coming from the measurement of inclusive \jpsi\ and $\psi(2S)$ rates
\cite{cdf_psi}, relies on two assumptions: (i) that all of the $\psi(2S)$ come
from $B$ decays, and (ii) that the \jpsi\ fraction $f_B$ is known.
The right hand side of Table~\ref{tpsicdf} -- which represents the
choice of parameters which comes closer to representing the CDF \jpsi\ spectrum
-- suggests a value for $f_B$ which is significantly smaller than the central
value used by CDF (namely 37\% vs. 63$\pm$ 17\% for $\pt(\psi)>6$ GeV
\cite{cdf_dpf}).  This
would decrease the effective \b\ cross section by a factor of 50\%. In
addition, the new processes responsible for the large $K$ factor apparent in
\jpsi\ production might affect $\psi(2S)$ production and could provide
enough rate to reduce the \b\ rate extracted from the assumption that $B$
decays are the only source of $\psi(2S)$.

Notice that $f_B$ could in principle be extracted
experimentally, for example by separating the direct \jpsi's from those due to
$B$ decays via the observation of the displaced vertex from which the $\psi$
orginates -- due to the long $B$ lifetime. UA1 measured $f_B$ by assuming that
direct \jpsi's are isolated while \jpsi's from $B$ decays are not, and studying
the isolation of the \jpsi's in the data.
This
assumption however might not be correct if other production mechanisms were
responsible for direct quarkonium production, such as for example gluon $\to$
\jpsi\ fragmentation \cite{braaten}.

It is very reasonable to expect that at some value of \pt\ the dominant
production mechanism for charmonium states will indeed be via gluon
fragmentation. The main reason being that direct production as described by the
LO mechanisms inhibits production at large \pt\ via a form factor suppression
(the probability that a charmonium bound state will hold together when
produced {\em directly} in an interaction with a large virtuality scale is
highly suppressed).
The fragmentation functions for the creation of $S$-wave charmonium ($\eta_c$
and \jpsi) in a gluon shower have recently been calculated \cite{braaten}\ and
work on the creation of $P$-wave states ($\chi$) is in progress
\cite{braaten2}.  It will be interesting to use these calculations in
order to extract the fragmentation contribution to charmonium production in the
regions of \pt\ explored experimentally, and verify whether these new
processes can account at least in part for the large observed $K$ factor.
The experimental detection of non-isolated \jpsi's from a primary vertex
-- and therefore presumably not coming from $B$ decays -- could provide a
strong indication that these processes are indeed present.

Similar measurements of the decay-vertex position of the
\psitwos\ would provide evidence in favour or against the current belief that
most of them come from $B$ decays. Once again the gluon
fragmentation contribution to production of this charmonium state could turn
out to be significant, and would manifest itself with a signal of non-isolated
prompt \psitwos.

Similarly interesting would be a separate measurement of the $\chi$ \pt\
spectrum, which is expected to be dominated by direct production rather than
$B$ decays. A preliminary measurement by CDF \cite{cdf_dpf} reports
$BR(\psi\to \mu^+\mu^-) \times$  $\sigma(\chi_c\to\psi\gamma;\; \pt_\chi > 7
GeV; \; \vert \eta \vert<0.5)$ = $3.2\pm0.3\pm1.2$ nb. Both $\chi_1$ and
$\chi_2$ are here included.  This can be compared with the range  $0.64 nb <
\sigma < 5.1 nb$ obtained using the calculation described in the previous
sections and the two extreme choices $(\muf,\lqcd)=(\mtr,215MeV)$ and
$(\muf,\lqcd)=(\mtr/4,275MeV)$. Using the above cross section and using the
inclusive $B\to \chi_{c1}$ branching ratio of $0.54\pm0.21$\% \cite{bchibr}, we
estimate that only a fraction of the order of 10\% or less -- depending on \pt
-- of the $\chi$'s come from $B$ decays. Since the production mechanisms for
$\chi_1$ and $\chi_2$ are different even at LO \cite{onia}, a separate
measurement of the two states would be welcome, even though their closeness in
mass makes it very hard to separate one from the other in practice.

It would also be interesting to evaluate the effects of resumming some of the
leading and next-to-leading corrections to the evolution of the initial state,
using the calculation of the $gg\to$ color-singlet NLO form factor calculated
in ref.\cite{32}.

An experimental measurement of the production cross section and \pt\ spectrum
for $\Upsilon$ states would be very useful in understanding the
quarkonium production mechanisms \cite{baier}. In this case, in fact, one would
have at least three advantages: (i) the masses involved are larger and
presumably both the non-relativistic approximation involved in the
determination of the quarkonium wave function and the QCD perturbative
expansion would work much more reliably than for charmonium; (ii) the signal
does not have a contamination similar to the one due to $B$ decays; (iii) the
\pt\ spectrum could hopefully be extended to very small values of \pt, possibly
even to \pt=0, thanks to the large mass of the $\Upsilon$ and the rather large
momentum and easier detection of the decay muons.

{\renewcommand{\arraystretch}{1.5}
\begin{table}
\begin{center}
\begin{tabular}{|c|c|c|} \hline
$\vert R_{1S}(0) \vert^2$ & $\vert R_{2S}(0) \vert^2$ &
$\vert R'_{1P}(0) \vert^2$
\\   \hline
0.7   &  0.4 &  0.006 \\
\hline
\end{tabular}
\caption{ \label{tparam}
\rightskip=1pc\leftskip=1pc\baselineskip=12pt
Values of wave functions used in the evaluation of the charmonium cross
sections.}
\end{center}
\end{table} }

{\renewcommand{\arraystretch}{1.5}
\begin{table}
\begin{center}
\begin{tabular}{|c|c|} \hline
$BR(B\to\psi X)\times BR(\psi\to\mu^+\mu^-)$ &
$BR(B\to\psi(2S) X)\times BR(\psi\to\mu^+\mu^-)$
\\   \hline
$7.7\times 10^{-4}$ & $3.6\times 10^{-5}$ \\
\hline
\end{tabular}
\caption{ \label{tbr}
\rightskip=1pc\leftskip=1pc\baselineskip=12pt
Values of BR's used in the evaluation of the \jpsi\ cross
sections.}
\end{center}
\end{table} }

{\renewcommand{\arraystretch}{1.2}
\begin{table}
\begin{center}
\begin{tabular}{|c||r|r|r|r|} \hline
$p_t^{min}$ & \multicolumn{2}{c|}{MRSD0} & \multicolumn{2}{c|}{MRSD--} \\
(GeV)       & $\Lambda_0$ & $\Lambda_+$  & $\Lambda_0$ & $\Lambda_+$  \\
\hline
 0 & 1.14E+04 & 1.34E+04 & 1.33E+04 & 1.57E+04  \\
 5 & 4.50E+03 & 5.22E+03 & 4.50E+03 & 5.27E+03  \\
10 & 1.05E+03 & 1.22E+03 & 9.45E+02 & 1.09E+03  \\
15 & 3.16E+02 & 3.56E+02 & 2.70E+02 & 3.11E+02  \\
20 & 1.15E+02 & 1.32E+02 & 9.69E+01 & 1.12E+02  \\
25 & 4.97E+01 & 5.58E+01 & 4.20E+01 & 4.75E+01  \\
30 & 2.40E+01 & 2.72E+01 & 2.06E+01 & 2.34E+01  \\
40 & 6.94E+00 & 7.85E+00 & 6.05E+00 & 6.80E+00  \\
50 & 2.60E+00 & 2.79E+00 & 2.16E+00 & 2.38E+00  \\
59 & 1.21E+00 & 1.30E+00 & 1.03E+00 & 1.14E+00  \\
\hline
\end{tabular}
\caption{ \label{tbcdf1}
\rightskip=1pc\leftskip=1pc\baselineskip=12pt
Integrated bottom quark $p_t$ distribution at 1.8 TeV. $m_b$=4.75 GeV,
\muf=\muo, $\Lambda_0=215$ MeV, $\Lambda_+=275$ MeV.}
\end{center}
\end{table} }

{\renewcommand{\arraystretch}{1.2}
\begin{table}
\begin{center}
\begin{tabular}{|c||r|r|r|r|} \hline
$p_t^{min}$ & \multicolumn{2}{c|}{MRSD0} & \multicolumn{2}{c|}{MRSD--} \\
(GeV)       & $\Lambda_0$ & $\Lambda_+$  & $\Lambda_0$ & $\Lambda_+$  \\
\hline
 0 & 2.17E+04 & 3.03E+04 & 2.69E+04 & 3.71E+04  \\
 5 & 7.23E+03 & 9.39E+03 & 7.68E+03 & 1.00E+04  \\
10 & 1.83E+03 & 2.27E+03 & 1.66E+03 & 2.09E+03  \\
15 & 5.78E+02 & 7.11E+02 & 4.93E+02 & 6.10E+02  \\
20 & 2.23E+02 & 2.68E+02 & 1.83E+02 & 2.25E+02  \\
25 & 9.86E+01 & 1.17E+02 & 7.98E+01 & 9.75E+01  \\
30 & 4.84E+01 & 5.63E+01 & 3.89E+01 & 4.72E+01  \\
40 & 1.44E+01 & 1.65E+01 & 1.23E+01 & 1.38E+01  \\
50 & 5.28E+00 & 6.04E+00 & 4.48E+00 & 4.87E+00  \\
59 & 2.23E+00 & 2.49E+00 & 2.04E+00 & 2.27E+00  \\
\hline
\end{tabular}
\caption{ \label{tbcdf2}
\rightskip=1pc\leftskip=1pc\baselineskip=12pt
Integrated bottom quark $p_t$ distribution at 1.8 TeV. $m_b$=4.75 GeV,
\muf=\muo/4, $\Lambda_0=215$ MeV, $\Lambda_+=275$ MeV.}
\end{center}
\end{table} }

{\renewcommand{\arraystretch}{1.2}
\begin{table}
\begin{center}
\begin{tabular}{|c||r|r|r|r|} \hline
$p_t^{min}$ & \multicolumn{2}{c|}{$\Lambda_0$} & \multicolumn{2}{c|}
{$\Lambda_+$} \\
(GeV)       & \mb=4.5 GeV & \mb=4.75 GeV & \mb=4.5 GeV & \mb=4.75 GeV \\
\hline
 0 & 2.6E+04 & 2.2E+04 & 3.8E+04 & 3.0E+04 \\
 5 & 8.0E+03 & 7.2E+03 & 1.0E+04 & 9.4E+03 \\
10 & 1.9E+03 & 1.8E+03 & 2.4E+03 & 2.3E+03 \\
20 & 2.3E+02 & 2.2E+02 & 2.8E+02 & 2.7E+02 \\
\hline
\end{tabular}
\caption{ \label{tbmass}
\rightskip=1pc\leftskip=1pc\baselineskip=12pt
Mass dependence of the integrated bottom quark $p_t$ distribution at 1.8 TeV.
MRSD0 parton
distributions, \muf=\muo/4, $\Lambda_0=215$ MeV, $\Lambda_+=275$ MeV.}
\end{center}
\end{table} }

{\renewcommand{\arraystretch}{1.2}
\begin{table}
\begin{center}
\begin{tabular}{|c||r|r|r|r|} \hline
$p_t^{min}$ & \multicolumn{2}{c|}{MRSD0} & \multicolumn{2}{c|}{MRSD--} \\
(GeV)       & $\Lambda_0$ & $\Lambda_+$  & $\Lambda_0$ & $\Lambda_+$  \\
\hline
 0 & 6.016E+03 & 7.110E+03 & 5.584E+03 & 6.606E+03  \\
 5 & 1.838E+03 & 2.149E+03 & 1.582E+03 & 1.848E+03  \\
10 & 2.953E+02 & 3.390E+02 & 2.482E+02 & 2.857E+02  \\
15 & 6.353E+01 & 7.183E+01 & 5.378E+01 & 6.188E+01  \\
20 & 1.776E+01 & 2.009E+01 & 1.566E+01 & 1.778E+01  \\
25 & 6.148E+00 & 6.714E+00 & 5.271E+00 & 6.110E+00  \\
30 & 2.419E+00 & 2.620E+00 & 2.143E+00 & 2.378E+00  \\
40 & 4.654E--01 & 5.086E--01 & 4.616E--01 & 5.098E--01  \\
50 & 1.312E--01 & 1.480E--01 & 1.324E--01 & 1.381E--01  \\
59 & 4.672E--02 & 5.346E--02 & 4.242E--02 & 4.552E--02  \\
\hline
\end{tabular}
\caption{ \label{tbua11}
\rightskip=1pc\leftskip=1pc\baselineskip=12pt
Integrated bottom quark $p_t$ distribution at UA1. $m_b$=4.75 GeV,
\muf=\muo, $\Lambda_0=215$ MeV, $\Lambda_+=275$ MeV.}
\end{center}
\end{table} }

{\renewcommand{\arraystretch}{1.2}
\begin{table}
\begin{center}
\begin{tabular}{|c||r|r|r|r|} \hline
$p_t^{min}$ & \multicolumn{2}{c|}{MRSD0} & \multicolumn{2}{c|}{MRSD--} \\
(GeV)       & $\Lambda_0$ & $\Lambda_+$  & $\Lambda_0$ & $\Lambda_+$  \\
\hline
 0 & 1.321E+04 & 1.833E+04 & 1.201E+04 & 1.647E+04  \\
 5 & 3.619E+03 & 4.705E+03 & 3.069E+03 & 3.985E+03  \\
10 & 6.036E+02 & 7.479E+02 & 4.986E+02 & 6.202E+02  \\
15 & 1.310E+02 & 1.600E+02 & 1.123E+02 & 1.325E+02  \\
20 & 3.635E+01 & 4.344E+01 & 3.193E+01 & 3.666E+01  \\
25 & 1.184E+01 & 1.391E+01 & 1.092E+01 & 1.235E+01  \\
30 & 4.452E+00 & 5.139E+00 & 4.210E+00 & 4.575E+00  \\
40 & 8.034E--01 & 9.621E--01 & 7.467E--01 & 8.312E--01  \\
50 & 1.723E--01 & 1.828E--01 & 1.765E--01 & 1.579E--01  \\
59 & 4.111E--02 & 4.241E--02 & 4.878E--02 & 4.510E--02  \\
\hline
\end{tabular}
\caption{  \label{tbua12}
\rightskip=1pc\leftskip=1pc\baselineskip=12pt
Integrated bottom quark $p_t$ distribution at 630 GeV. $m_b$=4.75 GeV,
\muf=\muo/4, $\Lambda_0=215$ MeV, $\Lambda_+=275$ MeV.}
\end{center}
\end{table} }

{\renewcommand{\arraystretch}{1.2}
\begin{table}
\begin{center}
\begin{tabular}{|c||r|r|r|r|r|r|r|} \hline
$p_t^{min,\psi} $ &
\multicolumn{3}{c|}{$\Lambda_0$ , \muf=\muo} &
\multicolumn{3}{c|}{$\Lambda_+$, \muf=\muo/4} \\
(GeV)
& $\sigma_B \cdot $ BR (nb)  &  $\sigma_\chi \cdot $ BR (nb)   &  $f_B$ (\%)
& $\sigma_B \cdot $ BR (nb)  &  $\sigma_\chi \cdot $ BR (nb)   &  $f_B$ (\%) \\
\hline
\hline
 3 &  2.6E+00 &  5.4E+00 & 32 &  5.4E+00 &  3.9E+01 & 12  \\
 4 &  1.7E+00 &  1.9E+00 & 46 &  3.4E+00 &  1.5E+01 & 19  \\
 5 &  1.1E+00 &  7.6E--01 & 58 &  2.2E+00 &  6.0E+00 & 26  \\
 6 &  6.7E--01 &  3.4E--01 & 66 &  1.4E+00 &  2.8E+00 & 33  \\
 8 &  2.9E--01 &  8.7E--02 & 77 &  6.3E--01 &  7.7E--01 & 45  \\
10 &  1.4E--01 &  2.9E--02 & 83 &  3.2E--01 &  2.6E--01 & 55  \\
12 &  7.8E--02 &  1.2E--02 & 86 &  1.7E--01 &  1.0E--01 & 62  \\
14 &  4.4E--02 &  4.9E--03 & 89 &  1.0E--01 &  5.0E--02 & 67  \\
16 &  2.6E--02 &  2.5E--03 & 91 &  6.1E--02 &  2.5E--02 & 71  \\
18 &  1.7E--02 &  1.3E--03 & 92 &  3.9E--02 &  1.3E--02 & 75  \\
20 &  1.1E--02 &  7.3E--04 & 93 &  2.5E--02 &  6.9E--03 & 78  \\
25 &  4.2E--03 &  1.9E--04 & 95 &  9.4E--03 &  1.9E--03 & 82  \\
30 &  1.9E--03 &  4.9E--05 & 97 &  4.1E--03 &  5.2E--04 & 88  \\
\hline
\end{tabular}
\caption{ \label{tpsicdf}
\rightskip=1pc\leftskip=1pc\baselineskip=12pt
Integrated $\psi$ \pt\ distribution from $B$ decays, from charmonium production
($\chi+\psi$) and relative $B$ fraction at 1.8 TeV. MRSD0, $\Lambda_0=215$ MeV,
$\Lambda_+=275$ MeV. BR($\jpsi\to \mu^+\mu^-$) included. }
\end{center}
\end{table} }

{\renewcommand{\arraystretch}{1.2}
\begin{table}
\begin{center}
\begin{tabular}{|c||r|r|r|r|r|r|r|} \hline
$p_t^{min,\psi} $ &
\multicolumn{3}{c|}{$\Lambda_0$ , \muf=\muo} &
\multicolumn{3}{c|}{$\Lambda_+$, \muf=\muo/4} \\
(GeV)
& $\sigma_B \cdot $ BR (nb)  &  $\sigma_\chi \cdot $ BR (nb)   &  $f_B$ (\%)
& $\sigma_B \cdot $ BR (nb)  &  $\sigma_\chi \cdot $ BR (nb)   &  $f_B$ (\%) \\
\hline
\hline
 3 &  3.0E+00 &  3.4E+00 & 46 &  7.8E+00 &  2.9E+01 & 21  \\
 4 &  1.6E+00 &  1.2E+00 & 56 &  4.0E+00 &  1.1E+01 & 27  \\
 5 &  8.3E--01 &  4.7E--01 & 63 &  2.1E+00 &  4.3E+00 & 33  \\
 6 &  4.6E--01 &  2.1E--01 & 69 &  1.2E+00 &  1.9E+00 & 37  \\
 8 &  1.6E--01 &  5.1E--02 & 75 &  4.0E--01 &  5.1E--01 & 43  \\
10 &  6.1E--02 &  1.6E--02 & 79 &  1.6E--01 &  1.7E--01 & 47  \\
12 &  2.7E--02 &  6.2E--03 & 81 &  6.7E--02 &  6.7E--02 & 50  \\
14 &  1.3E--02 &  2.7E--03 & 83 &  3.1E--02 &  2.9E--02 & 51  \\
16 &  6.6E--03 &  1.2E--03 & 84 &  1.6E--02 &  1.4E--02 & 53  \\
18 &  3.6E--03 &  6.1E--04 & 85 &  8.4E--03 &  7.2E--03 & 53  \\
20 &  2.0E--03 &  3.2E--04 & 86 &  4.5E--03 &  3.9E--03 & 53  \\
25 &  5.8E--04 &  7.2E--05 & 88 &  1.2E--03 &  9.2E--04 & 55  \\
30 &  2.0E--04 &  1.8E--05 & 92 &  3.6E--04 &  2.3E--04 & 61  \\
\hline
\end{tabular}
\caption{  \label{tpsiua1}
\rightskip=1pc\leftskip=1pc\baselineskip=12pt
Integrated $\psi$ \pt\ distribution from $B$ decays, from charmonium production
($\chi+\psi$) and relative $B$ fraction at 630 GeV. MRSD0, $\Lambda_0=215$ MeV,
$\Lambda_+=275$ MeV. BR($\jpsi\to \mu^+\mu^-$) included.}
\end{center}
\end{table} }

\clearpage

\begin{figure}
\caption[]
{\label{fua1_bpt}
Integrated $b$ \pt\ distribution at 630 GeV: UA1 data \cite{ua1_b2}\ versus NLO
QCD. The \jpsi\ point assumes a $B$ fraction in the inclusive \jpsi\ sample of
31$\pm$12\% \cite{ua1_psi2}.}
\end{figure}
\vskip -1.5cm

\begin{figure}
\caption[]
{\label{fua1_Bmespt}
Integrated $B$ meson \pt\ distribution at 630 GeV: UA1 data versus NLO
QCD. }
\end{figure}
\vskip -1.5cm

\begin{figure}
\caption[]
{\label{fua1_psi}
Differential $J/\psi$ \pt\ distribution at 630 GeV: UA1 data versus
different QCD contributions, as shown in the legend.}
\end{figure}
\vskip -1.5cm

\begin{figure}
\caption[]
{\label{fcdf_bpt}
Integrated $b$ \pt\ distribution at 1.8 TeV: CDF data \cite{cdf_dpf}\ versus
NLO
QCD. The \jpsi\ point assumes a $B$ fraction in the inclusive \jpsi\ sample of
63$\pm$17\% \cite{cdf_dpf}.}
\end{figure}
\vskip -1.5cm

\begin{figure}
\caption[]
{\label{fcdf_psi}
Differential $J/\psi$ \pt\ distribution at 1.8 TeV: CDF data versus
different QCD contributions, as shown in the legend.}
\end{figure}
\vskip -1.5cm

\begin{figure}
\caption[]
{\label{fcdf_psi2pt}
Differential $\psi(2S)$ \pt\ distribution at 1.8 TeV: CDF data versus
total QCD. The different contributions from direct production and $B$ decays
are labeled as in the previous Figure.}
\end{figure}
\vskip -1.5cm

\begin{figure}
\caption[]
{\label{fxua1}
Contribution to the $b$ cross section above given \pt\ thresholds as a function
of the gluon momentum fraction $x$.}
\end{figure}
\vskip -1.5cm

\vfill

\begin{thebibliography}{99}
\def	\nuke	#1#2#3{{\sl Nucl. Phys.} {\bf B#1}  (#2), #3}
\def	\pl  	#1#2#3{{\sl Phys. Lett.} {\bf #1B}  (#2), #3}
\def	\prl  	#1#2#3{{\sl Phys. Rev. Lett.} {\bf #1}  (#2), #3}
\def	\pr  	#1#2#3{{\sl Phys. Rev.} {\bf #1}  (#2), #3}
\def	\prd  	#1#2#3{{\sl Phys. Rev.} {\bf D#1}  (#2), #3}
\def	\zeit	#1#2#3{{\sl Z. Phys.} {\bf C#1}  (#2), #3}
\def	\cmp 	#1#2#3{{\sl Comm. Math. Phys.} {\bf #1}  (#2), #3}
\bibitem{nason}
	P. Nason,
	`Heavy Quark Production',
	to appear
	in "Heavy Flavours", Advanced Series on Directions in High Energy
	Physics, ed. M~Lindner, AJ~Buras. Singapore: World
       	Scientific (in press).
\bibitem{ua1_psi1}
	C. Albajar et al., UA1 Coll., \pl{200}{1988}{380}.
\bibitem{ua1_b1}
	C. Albajar et al., UA1 Coll., \pl{213}{1988}{405}.
\bibitem{nde}
	P.~Nason, S.~Dawson and R.~K.~Ellis,
	\nuke{303}{1988}{607}; \\
	W.~Beenakker, H. Kuijf, W.L. van Neerven and J. Smith,
        \prd{40}{1989}{54}.
\bibitem{nde_pt}
	P.~Nason, S.~Dawson and R.~K.~Ellis,
	\nuke{327}{1988}{49 }; \\
	W.~Beenakker, W.L. van Neerven, R. Meng, G.A. Schuler and J. Smith,
        \nuke{351}{1991}{507}.
\bibitem{gms}
	E.W.N. Glover, A.D. Martin and W.J. Stirling,
	\zeit{38}{1988}{473};\\
	E.W.N. Glover, F. Halzen and A.D. Martin,
	\pl{185}{1987}{441}.
\bibitem{ua1_psi2}
	C. Albajar et al., UA1 Coll., \pl{256}{1991}{112}.
\bibitem{ua1_b2}
	C. Albajar et al., UA1 Coll., \pl{256}{1991}{121}.
\bibitem{cdf_psiK}
	F. Abe et al., CDF Coll., \prl{68}{1992}{3403}.
\bibitem{cdf_psi}
	F. Abe et al., CDF Coll., FERMILAB-PUB-92/236-E (1992).
\bibitem{cdf_dpf}
	S. Vejcik, CDF Coll., presented at the 1992 DPF Meeting of the American
	Phys. Soc., Fermilab, November 11-14 1992;\\
	C. Boswell, CDF Coll., presented at the 1992 DPF Meeting of the American
	Phys. Soc., Fermilab, November 11-14 1992; FNAL-CONF-92/347-E; \\
	T. Fuess, CDF Coll., presented at the 1992 DPF Meeting of the American
	Phys. Soc., Fermilab, November 11-14 1992; FNAL-CONF-92/336-E; \\
	B.T. Huffman, CDF Coll., presented at the 1992 DPF Meeting of the American
	Phys. Soc., Fermilab, November 11-14 1992; FNAL-CONF-92/337-E.
\bibitem{smallx1}
	J.C. Collins and R.K. Ellis,
	\nuke{360}{1991}{3}.
\bibitem{smallx2}
	S. Catani, M. Ciafaloni and F. Hautmann,
	\nuke{366}{1991}{135}.
\bibitem{smallx3}
	E.M. Levin, M.G. Ryskin and Yu.M. Shabelsky,
	\pl{260}{1991}{429}.
\bibitem{NMC}
	P.~Amaudruz, et al., (NMC), CERN Preprint CERN-PPE/92-124.
\bibitem{CCFR}
	S.R. Mishra, et al., (CCFR) Nevis Rep. NEVIS-1466 (1992);
	\zeit{53}{1992}{51}
\bibitem{newmrs}
	A. Martin, R. Roberts and J. Stirling,
	RAL-92-021, DTP/92/16 (1992).
\bibitem{CTEQ} J.~Botts, et al., (CTEQ),  MSUHEP-92-27, Fermilab-Pub-92/371
	(1992)
\bibitem{berger1}
	E.L. Berger, R. Meng and W.K. Tung,
	\prd{46}{1992}{1895}.
\bibitem{berger2}
	E.L. Berger, R. Meng and J. Qiu,
	ANL-HEP-CP-92-79.
\bibitem{onia}
	E.L. Berger and D. Jones,
	\prd{23}{1981}{1521};\\
	R. Baier and R. R\"uckl, \zeit{19}{1983}{251};\\
	B. Humpert, \pl{184}{1987}{105};\\
	R. Gastmans, W. Troost and T.T. Wu, \nuke{291}{1987}{731}.
\bibitem{btopsi}
	H.~Schroder, (ARGUS), presented at {\em $25^{th}$ Intern. Conf. on HEP},
	Singapore, Aug.2-8 (1990);\\
	W. Chen, PhD thesis, (CLEO), Purdue Univ., May 1990; \\
	V. Papadimitriou, (CDF), personal communication.
\bibitem{peterson}
    	C. Peterson, D. Schlatter, I. Schmitt and P. Zerwas,
	\pr{D27}{83}{105}.
\bibitem{cdf_jet_shape}
	F. Abe et al., CDF Coll., FERMILAB-PUB/92/167-E (1992); \\
	S. Ellis, Z. Kunszt and D. Soper, UW/PT-91-01 (1992).
\bibitem{lep}
	Z. Kunszt and P. Nason, in {\it Z Physics at LEP}, eds. G. Altarelli ,
	R.~Kleiss and C.~Verzegnassi, CERN Rep. 89-08 (1989).
\bibitem{webber}
	S~Catani, et al., \pl{263}{491}{263}.
\bibitem{bodwin}
	G. Bodwin, E. Braaten and G.P. Lepage,
	\prd{46}{1992}{R1914}.
\bibitem{hagiwara}
	K. Hagiwara, A.D. Martin and A.W. Peacock,
	\zeit{33}{1986}{135}, as quoted in \cite{gms}.
\bibitem{mnr}
	M. Mangano, P. Nason and G. Ridolfi,
	\nuke{373}{1992}{295}.
\bibitem{geiser}
	A. Geiser,
	presented at XXVII Rencontres de Moriond, Les Arcs, PITHA 92/19 (1992);
	PhD thesis. RWTH, Aachen (1992).
\bibitem{braaten}
	E. Braaten and T.C. Yuan,
	NUHEP-TH-92-23; UCD-92-25 (1992).
	K. Hagiwara, A.D. Martin and W.J. Stirling,
	\pl{267}{1991}{527}.
\bibitem{braaten2}
	E. Braaten and T.C. Yuan, work in progress.
\bibitem{bchibr}
	R.A. Poling, in {\em Joint International Symposium and Europhysics
	Conference on High Energy Physics}, ed. S. Hegarty et al. (World
	Scientific, 1992), p.546.
\bibitem{32}
	S. Catani, E. D'Emilio and L. Trentadue,
	\pl{211}{1988}{335}.
\bibitem{baier}
	R. Baier and R.~R\"uckl, Ref.\cite{onia}.
\end{thebibliography}
\end{document}